\documentclass{article}

\usepackage{arxiv}

\usepackage{tikz}
\usetikzlibrary{arrows.meta, positioning, calc}
\usepackage{comment} 
\usepackage{amsmath,amssymb,amsfonts}
\usepackage{algorithm}
\usepackage{algorithmic}
\usepackage{graphicx}
\usepackage{textcomp}
\usepackage{tabularx}
\usepackage{xcolor}
\usepackage{subfigure}
\usepackage[caption=false]{subfig}
\usepackage[%
    style=ieee,%
    maxnames=3,%
    minnames=2,%
    backend=biber,%
	sorting=none,%
	natbib=true,%
	defernumbers=true,%
    doi=false,%
]{biblatex}%
\DefineBibliographyStrings{english}
   {andothers = \mkbibemph{et al\adddot}}
\bibliography{%
references.bib
}
\def\BibTeX{{\rm B\kern-.05em{\sc i\kern-.025em b}\kern-.08em
    T\kern-.1667em\lower.7ex\hbox{E}\kern-.125emX}}
\usepackage[nohyperlinks, printonlyused, nolist]{acronym}

\acrodef{iot}[IoT]{Internet of Things}
\acrodef{iiot}[IIoT]{Industrial Internet of Things}
\acrodef{mno}[MNO]{Mobile Network Operator}
\acrodef{urllc}[uRLLC]{Ultra Reliable and Low Latency Communication}
\acrodef{xurllc}[xURLLC]{extreme Ultra Reliable and Low Latency Communication}
\acrodef{cpps}[CPPS]{cyber-physical production systems}
\acrodef{qos}[QoS]{Quality of Service}
\acrodef{nm}[NetEm]{network emulation}
\acrodef{nin}[NiN]{Network-in-Network}
\acrodef{ncs}[NCS]{Networked Control System}
\acrodef{embb}[eMBB]{Enhanced Mobile Broadband}
\acrodef{kpis}[KPIs]{Key Performance Indicators}
\acrodef{dsm}[DSM]{Dynamic Spectrum Management}
\acrodef{cnc}[CNC]{Computerized Numerical Control}
\acrodef{fpga}[FPGA]{Field Programmable Gate Array}
\acrodef{cpf}[CPF]{control plane fabric}
\acrodef{dht}[DHT]{distributed hash table}
\acrodef{npn}[NPN]{Non-Public Network}
\acrodef{ai}[AI]{Artificial Intelligence}
\acrodef{rpi}[RPI]{Raspberry Pi}
\acrodef{sm}[SM]{spectrum manager}
\acrodef{sn}[SN]{sub-network}
\acrodef{snc}[SNC]{sub-network controller}
\acrodef{dt}[DT]{Digital Twin}
\acrodef{agv}[AGV]{Automated Guided Vehicle}
\acrodef{pn}[PN]{private network}
\acrodef{nin}[NiN]{Networks-in-Network}
\acrodef{mrat}[Multi-RAT]{multiple radio access technology}
\acrodef{ho}[HO]{Handover}
\acrodef{gnb}[gNB]{Next-Generation NodeB}
\acrodef{sgnb}[S-gNB]{Secondary Next-Generation NodeB}
\acrodef{ue}[UE]{User Equipment}
\acrodef{nnpn}[NNPN]{Nomadic Non-Public Network}
\acrodef{npn}[NPN]{Non-Public Network}
\acrodef{sb}[SB]{Spectrum Broker}
\acrodef{csm}[CSM]{Cognitive Spectrum Manager}
\acrodef{rrc}[RRC]{Radio Resource Control}
\acrodef{plmn}[PLMN]{Public Land Mobile Network}
\acrodef{mbsr}[MBSR]{Mobile Base Station Relay}
\acrodef{amf}[AMF]{Access and Mobility Management Function}
\acrodef{smf}[SMF]{Session Management Function}
\acrodef{upf}[UPF]{User Plane Function}
\acrodef{plmn}[PLMN]{Public Land Mobile Network}
\acrodef{mru}[MRU]{Mobile Registration Update}
\acrodef{ran}[RAN]{Radio Access Network}
\acrodef{cn}[CN]{Core Network}
\acrodef{du}[DU]{Distributed Unit}
\acrodef{rsrp}[RSRP]{Reference Signal Received Power}
\acrodef{sinr}[SINR]{Signal-to-Interference-plus-Noise Ratio}

\begin{document}
\title{Atomic Handover for 6G Nomadic Non-Public Networks Using Edge-Based Spectrum Brokering}

\author{
{Daniel~Lindenschmitt}\\
	Institute for Wireless Communication \\and Navigation\\
	RPTU Kaiserslautern-Landau\\
	\texttt{daniel.lindenschmitt@rptu.de} \\
	\And
	{Hans D.~Schotten}\\
	Institute for Wireless Communication \\and Navigation\\
	RPTU Kaiserslautern-Landau\\
	\texttt{schotten@rptu.de} \
}

\maketitle

\begin{abstract}
Nomadic Non-Public Networks (NNPN) are expected to play an important role in future 6G systems by enabling mobile and rapidly deployable network infrastructures for scenarios such as emergency response or temporary events. In such environments, maintaining seamless connectivity is challenging, as both network attachment and spectrum access may need to be adapted simultaneously when moving across heterogeneous infrastructures. In this paper, we investigate handover mechanisms for NNPN and propose a zero-touch approach that jointly considers mobility management and dynamic spectrum coordination. The proposed architecture introduces an edge-based Spectrum Broker in combination with a Cognitive Spectrum Manager to support an atomic handover procedure, where network selection and spectrum allocation are performed in a single step. The concept is evaluated using a MATLAB-based simulation of a mobile healthcare scenario, where an ambulance with its NNPN transitions between Public Land Mobile Networks (PLMN) and Non-Public Networks (NPN).
\end{abstract}

\keywords{
6G, Nomadic Network, Spectrum-Sharing, Handover
}
\section{Introduction}
Wireless communication systems have progressed through several generations with the objective of increasing network capacity, improving latency performance, and enabling connectivity for an expanding set of digital services. With the introduction of \acp{npn} in the 5G ecosystem, communication infrastructures can now be deployed to serve the operational needs of enterprises, industrial facilities, and public safety organizations \cite{npn}. 
Current \ac{npn} infrastructures are designed for stationary environments and rely on non-moving components such as \acp{ran} or \acp{cn}. While such architectures perform well in controlled locations, they are less suitable for applications in which connectivity must support mobile systems or temporary operations. Situations such as emergency response missions, mobile healthcare platforms, transportation systems, or temporary industrial sites require network infrastructures that remain operational while their physical location changes \cite{6GNomad_CSCN, miab_1}. \acp{nnpn} extend this concept by enabling network components to move together with the systems they support. 
This interaction introduces additional requirements for mobility management, particularly for the design of handover procedures to avoid interference between these networks.

A representative scenario is an ambulance equipped with a mobile base station that provides connectivity for onboard medical devices while maintaining links to external networks. During operation, the system encounters multiple network domains with different coverage characteristics. It must therefore transition between heterogeneous communication systems while preserving service continuity and meeting \ac{qos} requirements for critical applications. Two aspects complicate this process. First, the mobile base station fulfills a dual role: it operates as a \ac{gnb} for internal devices and as an \ac{ue} toward potential third-party networks. Second, it must adapt its spectrum usage when entering environments with different regulatory or sharing constraints \cite{10026262, 10754225}.

While further introducing the concept of \acp{nnpn} in Section~\ref{nomadic_nets}, the key contribution of this work is described in Section~\ref{spectrum_broker}, which is the integration of spectrum allocation and mobility management into an atomic handover mechanism coordinated by an edge-based \ac{sb} and a \ac{csm} for local spectrum decisions. Unlike conventional approaches that treat these processes separately, the proposed method performs joint decision making and execution within a unified control procedure. This reduces signaling overhead and enables faster adaptation in heterogeneous environments. We evaluate the approach using a simulation of a mobile healthcare scenario based on standardized radio channel models in Section~\ref{eval}. The results demonstrate that the proposed framework maintains connectivity across heterogeneous infrastructures while reducing signaling overhead and improving handover performance compared to conventional approaches, requiring only minor extensions to existing 3GPP signaling procedures.

\section{Related Work}
\label{rel_work}
Mobility management ensures that communication services remain available when network entities change their point of attachment to the radio infrastructure. In 5G systems, handover is typically controlled by the network and supported by measurements from the \ac{ue} \cite{miab_2}. The process follows several steps, starting with measurement configuration and reporting, followed by decision making at the serving \ac{gnb}, preparation of resources at the target node, and execution via \ac{rrc} signaling \cite{haghrah2023survey}. 
For example, Xn-based handover enables direct signaling between neighboring \acp{gnb} and reduces reliance on the core network, whereas NG-based handover involves the core network when direct links are not available \cite{handover_3}. Recent work has aimed to improve reliability and reduce interruption time during handover. Conditional handover prepares potential target cells in advance so that the \ac{ue} can switch as soon as predefined conditions are met. Another approach, Dual Active Protocol Stack handover, allows the \ac{ue} to maintain connections to both source and target cells during the transition, which helps to limit service interruption.
Dense deployments increase the number of handover events and can lead to frequent switching between cells. In addition, communication in higher frequency bands, such as millimeter-wave, is more sensitive to blockage and rapid signal changes, which makes stable handover more difficult. 

Looking ahead, 6G systems will further increase system complexity by combining terrestrial and non-terrestrial networks and supporting a much larger number of connected devices \cite{10054381, TN-NTN}. At the same time, emerging concepts such as Organic Networks and \ac{nin} architectures aim to make networks more adaptive and decentralized \cite{Organic6G}. Within this setting, \acp{nnpn} extend private network deployments by introducing mobility at the network level. Unlike conventional \ac{npn} infrastructures, \acp{nnpn} must handle both user mobility and the movement of the network itself \cite{6GNomadNet_ArchChal}.
Another open issue concerns the coordination of spectrum resources in such mobile network domains. Although dynamic spectrum sharing and cognitive radio techniques have been studied extensively, their integration with mobility management in nomadic scenarios is still limited \cite{10026262, spectrum_1}. 
These observations highlight the need for approaches that combine spectrum coordination with handover procedures in a unified framework. Existing approaches such as conditional handover, as defined by 3GPP, improve reliability but do not integrate spectrum negotiation into the handover execution. In contrast, the proposed approach embeds spectrum coordination directly into the mobility procedure.

\section{Nomadic Networks and Handover in 6G}
\label{nomadic_nets}
\subsection{Nomadic Non-Public Networks for 6G}
\acp{nnpn} take the idea of \acp{npn} a step further by removing the assumption of fixed deployment of the cellular network components. While conventional \acp{npn} are usually tied to locations such as factories, campuses, or hospitals, \acp{nnpn} are designed to move together with the systems they serve. In practice, this means that the network itself no longer remains static but instead follows the operational environment, maintaining local connectivity while interacting with whatever infrastructure is available in its surroundings \cite{lindenschmitt2024nomadic, 11328824}. These surrounding systems may include \acp{plmn} as well as \acp{npn} or other \acp{nnpn}. Even in areas without any possible terrestrial backhauling available, a non-terrestrial backhaul-link could be used \cite{rihan2024unified}. In this sense, an \ac{nnpn} combines two properties that are rarely addressed together: the controlled and application-specific behavior of private networks and the flexibility of mobile systems. Such networks typically operate under their own \ac{plmn} identity and may rely on licensed, shared, or dynamically assigned spectrum. At the same time, they can be integrated into larger communication ecosystems as \acp{nin} domains without losing their local control \cite{10054381}. 

\subsection{Dual Mobility}

\begin{figure}[htbp]
\centering
\includegraphics[width=1.0\columnwidth]{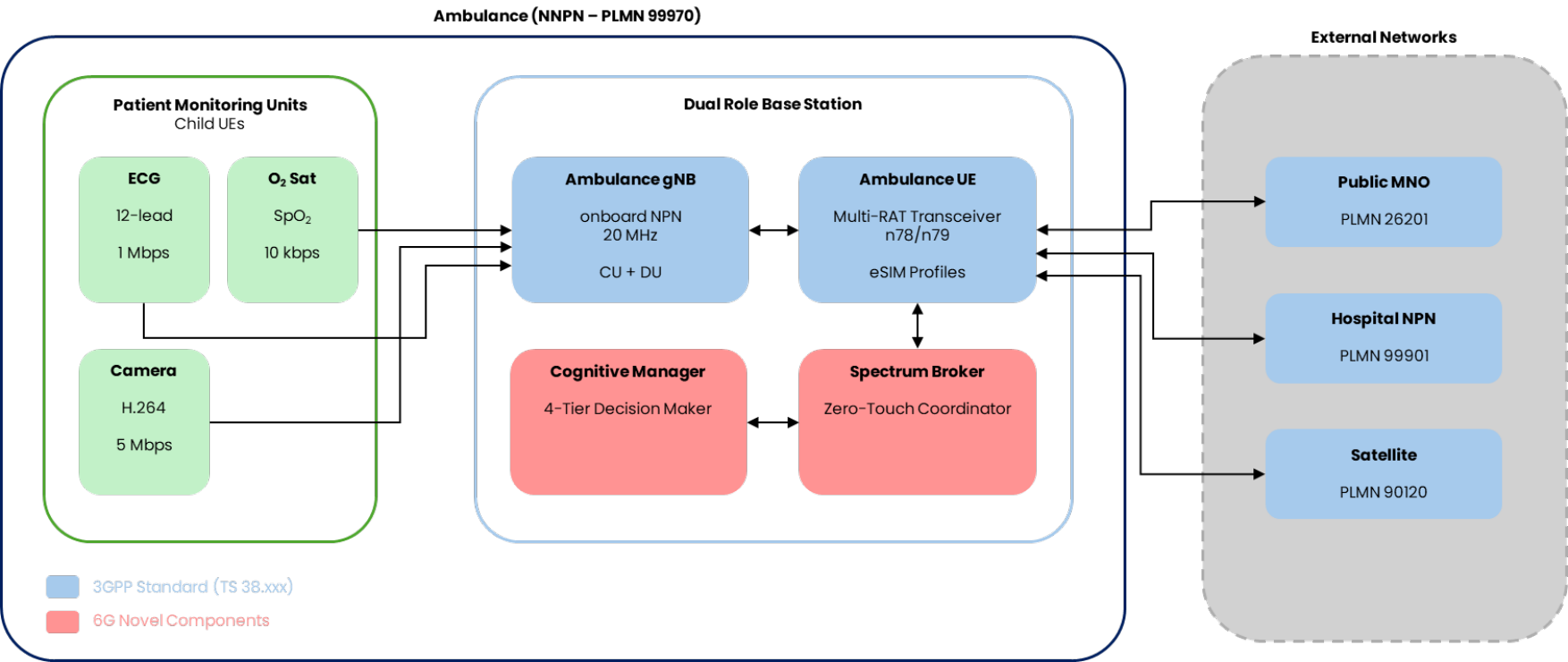}
\caption{Ambulance Use Case of a Dual Role Base Station}
\label{fig:usecase}
\end{figure}

A central feature of \acp{nnpn} is what can be described as dual mobility as shown in Figure~\ref{fig:usecase}. In traditional cellular systems, mobility management mainly deals with \acp{ue} moving relative to a  network. In \acp{nnpn}, this picture changes because the network itself is also in motion. As a result, mobility management has to deal with two processes at the same time. The first dimension corresponds to intra-\ac{npn} mobility, which describes the movement of devices within the coverage area of the nomadic network. In the ambulance scenario considered in this work, this includes medical monitoring devices such as ECG sensors, cameras, and diagnostic equipment connected to the onboard network. The second dimension is inter-\ac{npn} mobility, which refers to the movement of the entire network infrastructure relative to surrounding communication systems. As the ambulance moves through different geographic regions, the onboard network must transition between available communication infrastructures, including \acp{plmn}, \acp{npn} or non-terrestrial backhauling. Such transitions require coordination between heterogeneous network technologies and administrative domains. 
It is no longer sufficient to track signal quality alone. Instead, the system must ensure that the moving network and all connected devices remain continuously connected while transitioning between heterogeneous environments.
\begin{figure}[t]
\centering
\scalebox{0.7}{
\begin{tikzpicture}[
    every node/.style={font=\small},
    lifeline/.style={draw, thick},
    msg/.style={->, thick}
]

\def\xA{0}
\def\xB{3}
\def\xC{6}
\def\xD{9}

\node at (\xA,0.4) {UE};
\node at (\xB,0.4) {gNB (A)};
\node at (\xC,0.4) {gNB (B)};
\node at (\xD,0.4) {\ac{cn}};

\draw[lifeline] (\xA,0) -- (\xA,-6.4);
\draw[lifeline] (\xB,0) -- (\xB,-6.4);
\draw[lifeline] (\xC,0) -- (\xC,-6.4);
\draw[lifeline] (\xD,0) -- (\xD,-6.4);

\draw[msg] (\xB,-0.8) -- (\xC,-0.8) node[midway, above] {1. HO Request};

\draw[msg] (\xC,-1.5) -- (\xB,-1.5) node[midway, above] {2. HO Request ACK};

\draw[msg] (\xB,-2.2) -- (\xA,-2.2) node[midway, above] {3. HO Command};

\draw[msg] (\xA,-2.9) -- (\xC,-2.9) node[midway, above, xshift=3pt] {4. HO Confirm};

\draw[msg] (\xC,-3.6) -- (\xD,-3.6) node[midway, above] {5. HO Notify};

\draw[msg] (\xD,-4.3) -- (\xC,-4.3) node[midway, above] {6. HO ACK};

\draw[msg] (\xA,-5.0) -- (\xC,-5.0) node[midway, above] {7. \ac{mru}};

\draw[msg] (\xC,-5.7) -- (\xD,-5.7) node[midway, above] {8. \ac{mru}};

\draw[msg] (\xD,-6.4) -- (\xA,-6.4) node[midway, above] {9. Registration Accept};


\end{tikzpicture}
}
\caption{Canonical \ac{ho} based on 5G Release 16}
\label{fig:5G_HO}
\end{figure}
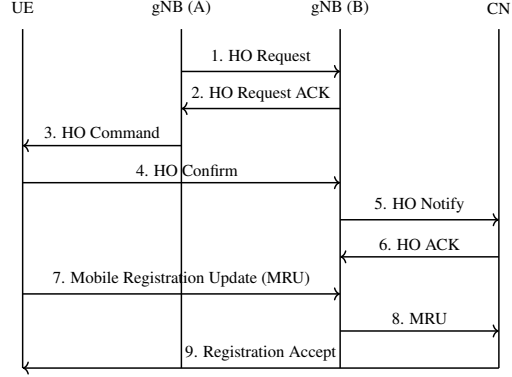
\subsection{Mobile Base Station Relay Architecture}
The canonical 5G handover standardized by 3GPP in Release 16 is illustrated in Figure~\ref{fig:5G_HO}. Recent developments in Release 18 have introduced the \ac{mbsr} concept to enhance mobile network infrastructures. The idea is to let a moving platform host network functions while staying connected to external networks. By combining relay capabilities with distributed base station elements, the \ac{mbsr} architecture makes \acp{nnpn} possible. An \ac{mbsr} generally consists of a mobile terminal, which keeps the backhaul link with external networks active and a \ac{gnb}, which handles radio access for devices inside the local nomadic network. In practice, an emergency medical vehicle can use this setup to connect medical equipment to the onboard \ac{gnb}, while the relay terminal links to external networks such as cellular systems or satellites. To the outside world, the vehicle appears as a single communication node. This makes it easier to manage connectivity for the internal devices, since only the relay needs to perform handovers with external networks. 

Ensuring uninterrupted service during network transitions is critical for \ac{nnpn} operations. This is particularly true in healthcare applications, where constant transmission of medical data can be a matter of safety. Research on nomadic networks points to several key requirements for mobility management: high availability, low handover delays, predictable handling of mobility, and reliable links across different network types. Maintaining session continuity is especially important for real-time monitoring or interactive communication. Handover processes must reduce interruptions and keep active sessions stable. Meeting these goals requires mechanisms that bring together network selection, spectrum allocation, and resource management in a coordinated way. The next section presents the \ac{sb} architecture, designed to address these challenges.

\section{Spectrum Broker–Based Handover}
\label{spectrum_broker}
\subsection{Architecture Overview}
Mobility management in \acp{nnpn} differs fundamentally from conventional cellular deployments. In traditional systems, handover is triggered when a user device moves relative to fixed base stations. In nomadic scenarios, however, the access infrastructure itself may move together with the \acp{ue}. 
In addition to this mobility aspect, \acp{nnpn} operate in heterogeneous environments that include \acp{plmn} and \acp{npn}, each with different spectrum policies and administrative constraints. Mobility decisions must therefore consider both radio conditions and spectrum availability. To address this, we introduce a handover architecture that integrates spectrum coordination directly into the mobility decision process while remaining compatible with established 3GPP signaling procedures.

The proposed architecture extends the 5G system by introducing coordination functionality at the network edge, while core entities such as \ac{amf}, \ac{smf}, and \ac{upf} remain unchanged. A central component is the \ac{sb}, which is deployed within each \ac{nnpn}. It collects measurement data, evaluates candidate networks, and coordinates spectrum access when the network enters a new operational domain. In addition, a \ac{csm} evaluates local spectrum conditions and determines suitable frequency configurations. By placing these functions at the edge, decisions can be taken locally, which reduces latency during transitions. 

The atomic procedure embeds spectrum parameters into \ac{rrc} reconfiguration messages, avoiding post-handover spectrum adaptation and reducing signaling exchanges. All \ac{sb}-related decisions are performed locally at the edge, limiting additional latency to sub-10 ms and avoiding round-trip delays to centralized control entities. The latency estimate is based on local edge processing assumptions without core-network interaction.
\begin{figure}[t]
\centering
\resizebox{0.8\columnwidth}{!}{
\begin{tikzpicture}[
    font=\footnotesize,
    >=latex
]

\node at (0,0) (nnpn) {\ac{nnpn}};
\node at (3.5,0) (sb) {\ac{sb}/\ac{csm}};
\node at (7,0) (tgnb) {gNB (B)};
\node at (10.5,0) (net) {Target Network};

\draw (0,-0.3) -- (0,-5);
\draw (3.5,-0.3) -- (3.5,-5);
\draw (7,-0.3) -- (7,-5);
\draw (10.5,-0.3) -- (10.5,-5);


\draw[->] (0,-0.8) -- (3.5,-0.8);
\node at (1.75,-0.6) {1. Measurement Reports};

\draw[->] (3.5,-1.5) -- (7,-1.5);
\node at (5.25,-1.3) {2. HO + Spectrum Request};

\draw[->] (7,-2.2) -- (3.5,-2.2);
\node at (5.25,-2.0) {3. Resource + Spectrum Grant};

\draw[->] (3.5,-3.0) -- (0,-3.0);
\node at (1.75,-2.8) {4. Unified RRC Reconfiguration};

\draw[->] (0,-3.8) -- (7,-3.8);
\node at (3.5,-3.6) {5. HO Execution};

\draw[->] (7,-4.5) -- (10.5,-4.5);
\node at (8.75,-4.3) {6. HO Confirm};

\end{tikzpicture}
}
\caption{Atomic handover signaling with \ac{csm} via edge-based \ac{sb}}
\label{fig:signaling_flow}
\end{figure}
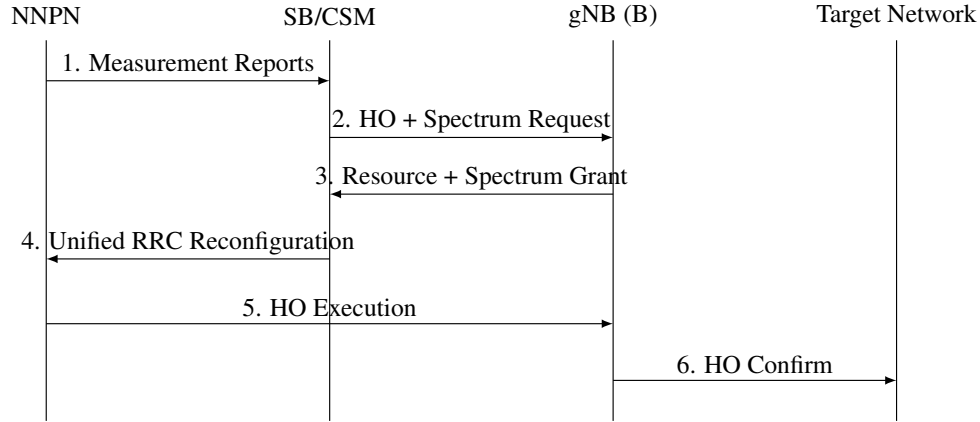

\subsection{Handover Operation and Spectrum Coordination}
Network transitions in nomadic scenarios may occur across infrastructures that either are operated by public \acp{mno}, the same private stakeholder as a \textit{cooperative \ac{npn}} such as the \ac{nnpn} or to a third-party private operator as an \textit{non-cooperative \ac{npn}}. When the transition occurs between two \acp{gnb} operated by the same stakeholder, the process corresponds to a horizontal handover and follows standard Xn-based procedures defined by the 3GPP framework. These procedures enable direct signaling between neighboring \acp{gnb} and allow efficient handover execution with minimal involvement of the \ac{cn}. In contrast, vertical handovers occur when the nomadic network connects to a \ac{gnb} operated by a different stakeholder, either a \ac{plmn} or a \ac{npn}, or a non-terrestrial backhaul-link. These transitions require additional evaluation because spectrum usage policies, frequency bands, and access conditions may differ between the involved infrastructures. 
Figure~\ref{fig:signaling_flow} abstracts lower-layer signaling to highlight the integration of the \ac{sb}, which supervises this process by continuously evaluating measurement reports obtained from the network environment. These reports include e.g., the \ac{rsrp} or interference indicators such as the SINR. When radio conditions indicate that a transition is beneficial, the broker evaluates whether the current spectrum configuration remains valid in the target environment. If the existing configuration cannot be maintained due to regulatory or technical constraints, the broker initiates a coordination process with the destination network. In the ambulance scenario considered in this work, this coordination enables the hospital network to temporarily adjust its spectrum allocation and provide a part of its bandwidth to the incoming \ac{nnpn}. This cooperative spectrum sharing allows the ambulance network to operate within the hospital domain while maintaining service continuity for critical applications.

Spectrum adaptation is realized through four operational strategies. In the \textit{default mode}, the system operates with a predefined configuration that maximizes available bandwidth under unconstrained conditions. In \textit{cognitive mode}, spectrum sensing is performed to detect interference and dynamically adapt bandwidth to avoid occupied frequencies. In \textit{negotiated mode}, the \ac{sb} coordinates with \textit{cooperative \acp{npn}} to obtain temporary spectrum access, e.g., by requesting a dedicated bandwidth part from a hospital network. Finally, in \textit{vacation mode}, when the primary band becomes unavailable, the system switches to an alternative frequency band, ensuring continued operation at reduced performance. Conventional mobility procedures treat network selection and spectrum configuration as separate steps, increasing signaling overhead and transition time. In contrast, the proposed approach integrates both into a single decision process. The \ac{sb} evaluates candidate networks based on radio conditions, spectrum availability, and expected service quality, and generates a unified configuration comprising both connection parameters and spectrum allocation following the Algorithm~\ref{alg:sb}. A weighted utility function as defined in Equation~\ref{eq:candidated_selection} is used to rank candidates, where \ac{rsrp} and spectrum feasibility $S_i$ are jointly optimized, the weights $w_1$, $w_2$ and $w_3$ are empirically chosen and normalized to sum to one. A detailed sensitivity analysis is left for future work.
\begin{equation}
U_i = w_1 \cdot \text{RSRP}_i + w_2 \cdot \text{SINR}_i + w_3 \cdot S_i
\label{eq:candidated_selection}
\end{equation}
The resulting configuration is transmitted through a \ac{rrc} reconfiguration message. This allows the system to connect to the target network while directly applying the required spectrum settings.

\begin{algorithm}
\caption{Spectrum Broker Decision Process}
\label{alg:sb}
\begin{algorithmic}[1]
\STATE Collect measurements and identify candidate networks
\FOR{each candidate network}
\STATE Evaluate signal quality and check spectrum availability 
\IF{spectrum conflict detected}
\STATE Initiate negotiation with target network
\ENDIF
\ENDFOR
\STATE Rank candidates based on \ac{qos} and spectrum feasibility
\STATE Select optimal network and spectrum configuration
\STATE Generate unified \ac{rrc} reconfiguration message
\STATE Execute atomic handover
\end{algorithmic}
\end{algorithm}

\subsection{Signaling and Complexity Considerations}
In contrast to conventional approaches, where spectrum configuration is applied after network attachment, the proposed method embeds spectrum parameters directly into the handover. This avoids intermediate reconfiguration steps and results in a more compact signaling procedure. The computational complexity of the decision process scales linearly with the number of candidate networks. Since all evaluations are performed at the network edge, latency remains bounded and independent of centralized control functions. The proposed approach provides an effective solution for mobility management in \ac{nnpn} scenarios while remaining compatible with existing 3GPP mechanisms.

\section{Evaluation}
\label{eval}
\subsection{Simulation Setup}
We evaluate the proposed handover and spectrum coordination approach using a simulation scenario that reflects a mobile healthcare network. Since such systems are not yet widely deployed, the analysis relies on a simulation environment that allows detailed investigation of mobility procedures and spectrum management behavior. The simulation is implemented in MATLAB using the 5G Toolbox. It reproduces the functionality of the \ac{nnpn}, including \ac{sb} and \ac{csm}, introduced in the previous section. The radio propagation follows the 3GPP TR 38.901 Urban Macro model, capturing path loss, shadowing, and penetration effects typical for urban environments. All reported results represent averages over multiple simulation runs with varying mobility conditions to ensure robustness against scenario-specific effects. The evaluation focuses on mobility robustness and control-plane efficiency during handover events. Metrics such as interruption time, signaling overhead, and packet loss are therefore prioritized, while throughput-related performance is left for future investigation, as the focus of this work is on control-plane efficiency and mobility robustness during handover events.
\begin{figure}[htbp]
\centering
\includegraphics[width=1.0\columnwidth]{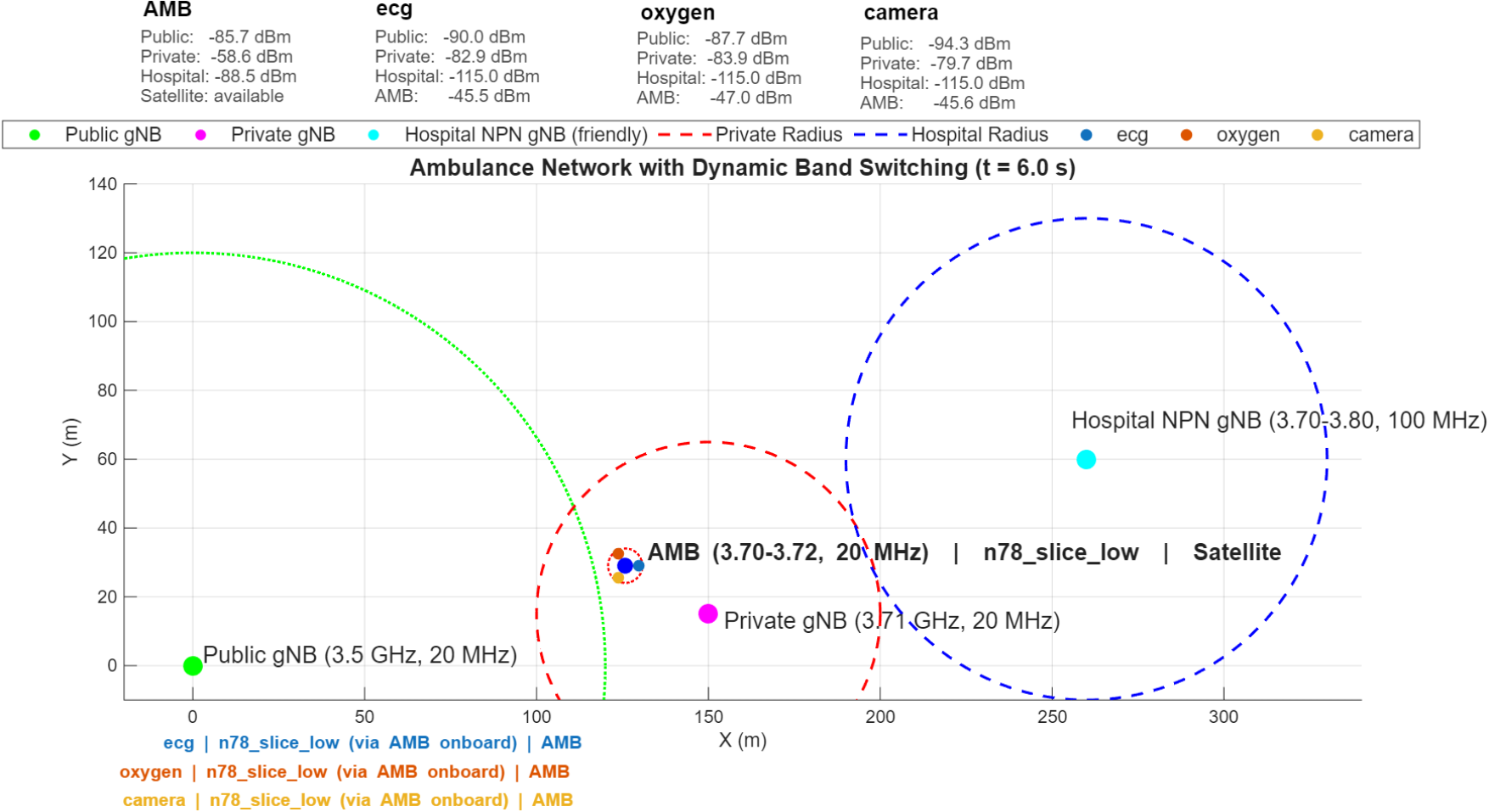}
\caption{Simulation overview}
\label{fig:overview}
\end{figure}
The system operates in the n78 band between 3.7 GHz and 3.8 GHz. During operation, the network continuously monitors its radio environment. Measurements from neighboring infrastructures are collected periodically and forwarded to the \ac{sb}. Based on these inputs, the broker determines whether the current connection remains suitable or if a handover based on RSRP with Time-to-Trigger should be initiated. To assess the advantages of the proposed architecture, we compare it with a baseline approach, as shown in Figure~\ref{fig:5G_HO}, in which network selection and spectrum allocation are executed sequentially. In this reference case, spectrum reconfiguration is applied only  after a successful network attachment, leading to additional signaling steps. The simulated scenario represents an ambulance equipped with a \ac{nnpn} moving through an urban environment as shown in Figure~\ref{fig:overview}. As the vehicle progresses, the onboard network encounters different infrastructures that vary in availability and signal quality. The simulation spans 18 seconds and is divided into 180 time steps. At each step, signal conditions are updated and the system evaluates whether a reconfiguration is required.

Initially, the ambulance connects to a terrestrial \ac{plmn}. As distance increases, signal strength degrades until it falls below a threshold, triggering a handover to a satellite backhaul, because no bandwidth could be negotiated with the \textit{non-cooperative \ac{npn}}. Later, the ambulance approaches a \textit{cooperative \ac{npn}}, in this case the hospital equipped with a \ac{npn}. Once this network becomes available, the \ac{sb} evaluates it as a candidate and initiates a transition. During this phase, the hospital network temporarily assigns part of its spectrum to the ambulance system. When the vehicle leaves the coverage area, this allocation is released. 

\subsection{Handover Analysis and Spectrum Broker Behavior}
\begin{figure}[htbp]
	\centering
    \subfigure[N2-based handover via \ac{amf}]{\includegraphics[width=0.9\columnwidth]{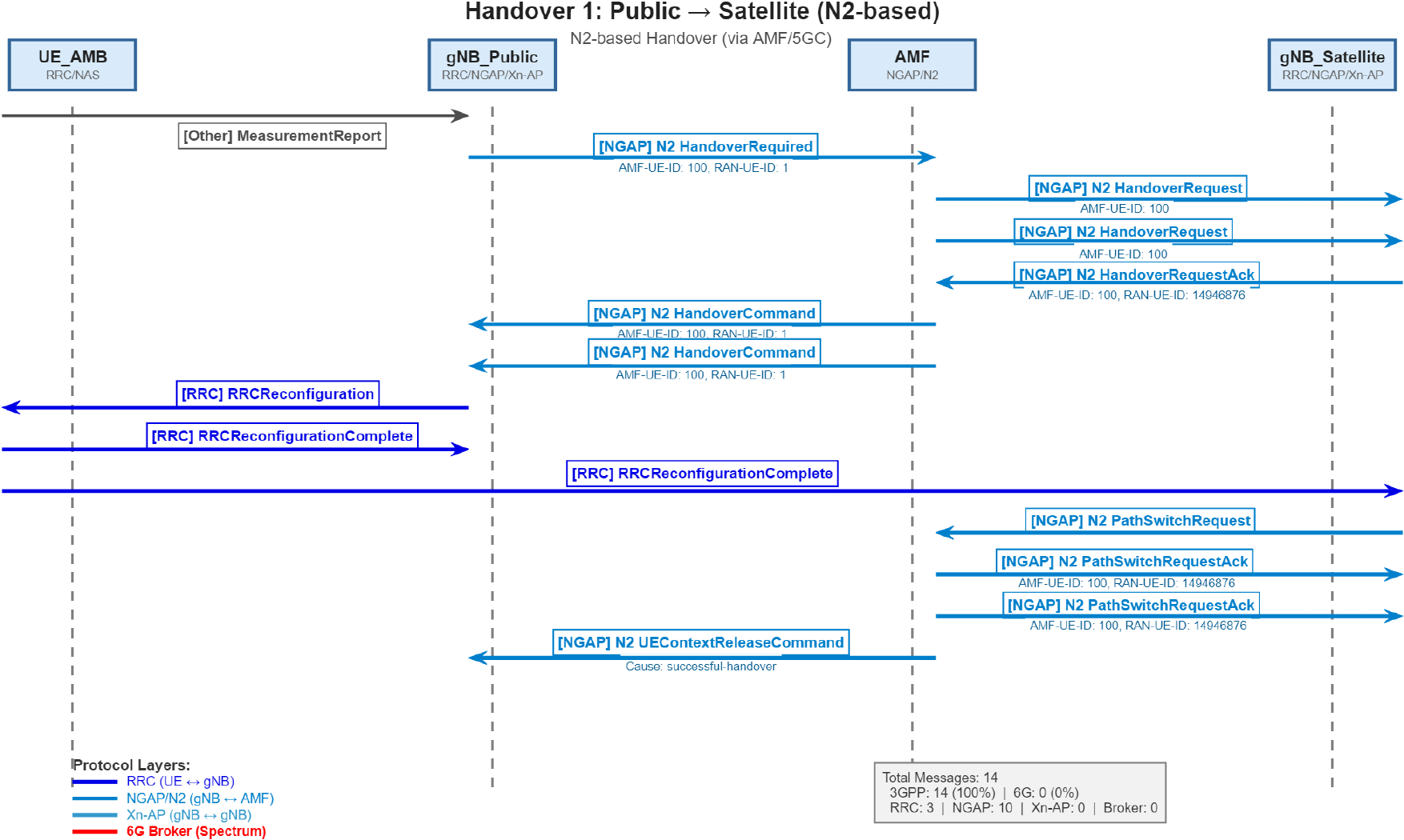}}\label{fig:handover_b}
    \subfigure[Xn-based handover via \ac{sb}]{\includegraphics[width=0.9\columnwidth]{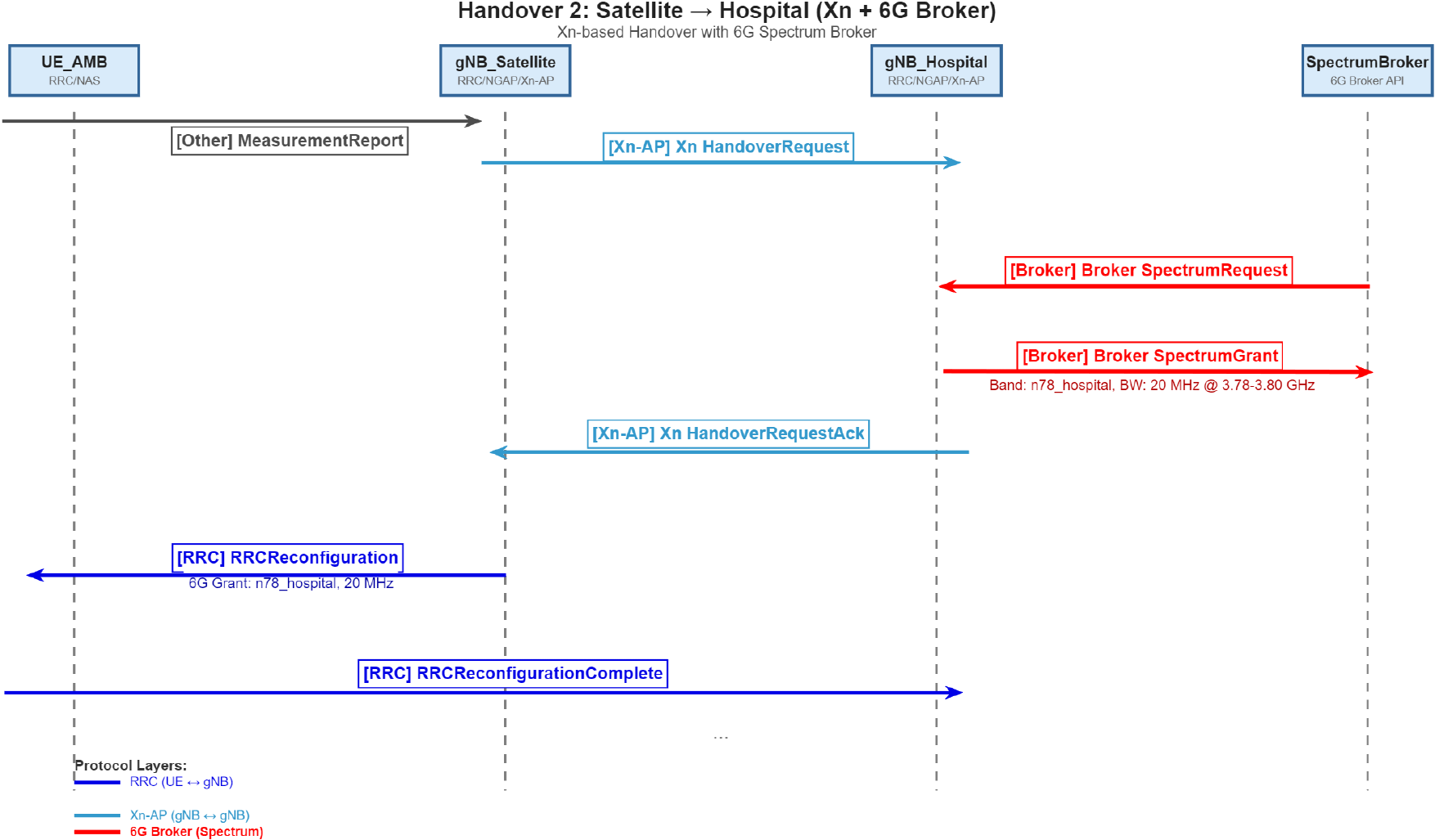}}\label{fig:handover_a}    
\caption{Handover behavior during movement of \ac{nnpn}}
\end{figure}
In Figure~5a the handover process from a non-terrestrial backhaul to a \textit{cooperative \ac{npn}} based on a Xn-based handover in combination with the \ac{sb} is visualized, while in Figure~5b a N2-based handover from the \textit{cooperative \ac{npn}} to a non-terrestrial backhaul is shown. The results show that the proposed architecture enables stable transitions across all involved infrastructures. Each handover is completed successfully, and ongoing sessions remain active throughout the scenario. The observed interruption time remains below 300~ms, meeting the design target. Additional simulations with varying mobility speeds and network conditions showed consistent performance trends, indicating that the proposed approach remains robust under different operational dynamics. This behavior results from the integrated handover mechanism, where network selection and spectrum configuration are executed jointly rather than sequentially. The use of a Time-to-Trigger parameter further improves stability. By requiring the handover condition to persist for a defined interval, the system avoids unnecessary transitions caused by short-term signal fluctuations. In addition to handover success and interruption time, we analyze the following metrics:
\begin{itemize}
\item \textbf{Signaling overhead:} reduced due to combined execution
\item \textbf{Spectrum usage:} improved through dynamic allocation
\item \textbf{Stability:} fewer handovers due to integrated decisions
\end{itemize}
The baseline based on Figure~\ref{fig:5G_HO} and the atomic approach introduced in Figure~\ref{fig:signaling_flow} are compared in Table~\ref{tab:handover_comparison}. The proposed approach reduces intermediate signaling steps and avoids separate spectrum reconfiguration phases, resulting in more consistent transitions and faster handover. The behavior of the \ac{sb} can be observed through the decisions of the \ac{csm}. Over the simulation, 180 spectrum allocation decisions are performed. Outside the hospital coverage area, the system operates in its \textit{default mode}, as no conflicting signals are present. When entering the hospital domain, the \ac{sb} detects the available infrastructure and initiates the \textit{negotiation mode}. This results in a temporary allocation of spectrum resources to the ambulance. After leaving the hospital area, the shared allocation is released and the system returns to its standard configuration. 

To complement the connectivity-based evaluation, a simple traffic model is introduced to assess packet-level performance during handover events. Packets are generated at fixed intervals and transmitted over the active connection. During handover transitions, packet delivery is monitored to evaluate loss and delay characteristics. The results in Table~\ref{tab:handover_comparison} show that the proposed atomic handover approach reduces packet loss during transitions from approximately 3.5\% in the baseline scenario to 1.8\%. This improvement is attributed to the elimination of intermediate spectrum reconfiguration steps and the reduced interruption time. While throughput is not explicitly evaluated, packet loss and interruption time provide indirect indicators of data-plane performance during handover. Reduced interruption duration and lower packet loss imply improved continuity of data transmission, particularly for delay-sensitive services. These findings indicate that integrating spectrum coordination into the handover process not only improves connectivity continuity but also enhances application-level performance for delay-sensitive services.
\begin{table}[htbp]
\centering
\caption{Comparison of Baseline and Atomic Handover}
\label{tab:handover_comparison}
\begin{tabular}{|l|c|c|}
\hline
\textbf{Metric} & \textbf{Baseline} & \textbf{Atomic} \\
\hline
Handover Interruption Time [ms] & 420 & 280 \\
\hline
Signaling Steps (per HO) & 6 & 4 \\
\hline
Spectrum Reconfiguration Delay [ms] & 150 & 0 (integrated) \\
\hline
Packet Loss During HO [\%] & 3.5 & 1.8 \\
\hline
Handover Success Rate [\%] & 96.2 & 98.7 \\
\hline
\end{tabular}
\end{table}
\subsection{Limitations}
The results indicate that integrating spectrum coordination with mobility management improves transition stability and reduces interruption time. The reduction in signaling steps contributes directly to faster and more reliable handovers. Furthermore, dynamic spectrum negotiation enables operation in environments with restricted resources. However, the evaluation remains partly limited. The scenario includes a single mobile network and a small number of handover events. In addition, no multi-user environment is considered and performance metrics such as throughput are not evaluated in detail. Future work will extend the simulation to more complex scenarios, including multiple users, longer durations, and additional performance indicators. This will allow a more comprehensive assessment of the proposed approach.

\section{Conclusion and Future Work}
\label{concl}
This work presented a handover architecture designed for 6G \acp{nnpn}. The architecture introduces an edge-based \ac{sb} together with a \ac{csm} that enable coordinated spectrum usage during handover procedures. By combining network selection and spectrum allocation within a single atomic decision process, the proposed approach reduces signaling overhead while remaining compatible with established 3GPP communication procedures. Simulation results based on a mobile healthcare scenario demonstrate that the architecture supports seamless transitions between satellite infrastructure, public networks, and private network domains while maintaining low decision latency. These results indicate that broker-assisted spectrum coordination can support autonomous connectivity management in future nomadic network deployments.  Future work will extend the evaluation toward more comprehensive performance analysis.

\section*{Acknowledgment}
The authors acknowledge the financial support by the German \textit{Federal Ministry of Research, Technology and Space (BMFTR)} within the project Open6GHub+ \{16KIS2406\} and X-COM \{16KISS007K\}.

{%
\printbibliography%
}%

\end{document}